\documentclass[twocolumn,showpacs,pra,raggedbottom]{revtex4}
\usepackage{graphicx}
\begin{document}
\setlength\arraycolsep{0pt}

\title{Feshbach resonance scattering under cylindrical harmonic
confinement}

\author{V. A. Yurovsky}

\affiliation{School of Chemistry, Tel Aviv University, 69978 Tel Aviv,
Israel}

\date{\today}

\begin{abstract}A problem of collisions of atoms with two-channel
 zero-range
interaction in an atomic waveguide is solved by using of a
 renormalization
procedure. A matching of the solution to a solution of the related
 one-dimensional
problem leads to relation between the one-dimensional and
 three-dimensional
scattering parameters. The scattering amplitude and bound states for
 the confined
system demonstrate differences from the related free and
 one-dimensional
systems.\end{abstract}

\maketitle    

\section*{Introduction}

Quasi-one-dimensional atomic gases have been realized recently in
elongated atomic traps (see Refs.\ \cite{S02,M02}), two-dimensional
optical lattices (see Ref.\ \cite{G01}), atomic waveguides (see Ref.\
\cite{L02} and references therein), and atomic integrated optics
 devices
(see Ref.\ \cite{S03} and references therein). These systems  attract
recently increased attention due to their possible applications to
atomic interferometry, quantum measurements, and  quantum
 computations.
Ultracold atoms under tight cylindrical confinement could reach the
``single-mode'', or quasi-one-dimensional regime, where only the
 ground
state of transverse motion is significantly populated at the thermal
equilibrium. An analysis of two-body collisions in this regime in
 Ref.\
\cite{O98} demonstrates that the center-of-mass motion can be
 separated
in a case of harmonic confinement and a zero-range interaction between
free atoms leads to an effective one-dimensional interaction between
confined atoms. The one-dimensional interaction strength demonstrates
resonant properties as a function of the ratio of the transverse width
and the elastic scattering length. This confinement induced resonance
has been analyzed in Refs.\ \cite{BMO03,MBO04}. A related problem of
 two
atoms under three-dimensional harmonic confinement has been considered
in Refs.\ \cite{BERW98,TWMJ00,BG02,BTJ02,BTJ03}. The works
\cite{O98,BMO03,MBO04,BERW98,TWMJ00,BG02,BTJ02,BTJ03} analyzed a case
 of
Bose atoms, describing their interactions by a Fermi pseudopotential.
 A
case of confined fermions has been considered in Refs.\
\cite{GB04,KB04,GO03,GO04}.

The previous works deal with a case of interatomic interaction
involving a single channel. However, a possibility of scattering
 properties
control due to the effect of Feshbach resonance (see Ref.\
 \cite{TTHK99})
brings attention to multichannel problems. A Feshbach resonance can
 appear
if the energy of collision in an open channel is close to the bound
 state
energy in a closed channel. A description of non-diagonal elements of
 a
potential matrix in multichannel problems by a Fermi pseudopotential
 seems
to be problematic due to its non-hermitian form. The present work,
following Ref.\ \cite{K02}, considers Hermitian $\delta $ function
 interactions.
Such interactions lead do divergences, requiring a renormalization of
scattering parameters. Probably, a multichannel zero-range potential
 method
of Ref.\ \cite{KZ03} could be used here as an alternative approach.

The work is outlined as follows. A renormalization procedure for
two-channel scattering under cylindrical harmonic confinement is
presented in Sec.\ \ref{secRenorm}. Section \ref{sec1D} describes two
models of two-channel one-dimensional systems: the atom-molecule one
and the two-state one. The results are discussed in Sec.\
 \ref{secRes}.
They include relation of one-dimensional scattering parameters to
three-dimensional ones, analyzes of scattering amplitudes and bound
states. A system of units in which Planck's constant is $\hbar =1$ is
 used
below.

\section{Renormalization\label{secRenorm}}

Following Ref.\ \cite{K02}, consider two-channel scattering of
atoms with $\delta $ function interactions under the external harmonic
confinement described by the potential $V_{\text{conf}}\left( {\bf
 r}\right) $. Close-coupled equations
for the wavefunction of the open channel $\psi _{a}\left( {\bf
 r}\right) $ and the amplitude for the
system to be in the closed channel $\psi _{m}$  have the form,
\begin{eqnarray}
E\psi _{a}\left( {\bf r}\right) &=&\left\lbrack -{1\over m}\nabla
 ^{2}+V_{a}\delta \left( {\bf r}\right) +V_{\text{conf}}\left( {\bf
 r}\right) \right\rbrack \psi _{a}\left( {\bf r}\right)  \nonumber
\\
&&+V_{am}\delta \left( {\bf r}\right) \psi _{m}\label{CC3D}
\\
E\psi _{m}&=&D_{3D}\psi _{m}+V^{*}_{am}\psi _{a}\left( 0\right)  .
 \nonumber
\end{eqnarray}
Here $m$ is the mass of the atom, $E$ and ${\bf r}$ are,
 respectively, the energy
and coordinate vector of the relative motion, $V_{a}$  is the
 strength of the
open channel potential, $V_{am}$  is the coupling strength, and
 $D_{3D}$  is a bound
state energy in the closed channel. All the energies here are counted
 from
the open channel threshold. Elimination of $\psi _{m}$  from Eqs.\
(\ref{CC3D})
leads to the one-channel Schroedinger equation
\begin{equation}
E\psi _{a}\left( {\bf r}\right) =\left\lbrack -{1\over m}\nabla ^{2}
+V_{\text{eff}}\left( E\right) \delta \left( {\bf r}\right)
+V_{\text{conf}}\left( {\bf r}\right) \right\rbrack \psi _{a}\left(
 {\bf r}\right)  \label{Eff3D}
\end{equation}
with the energy-dependent effective interaction strength
\begin{equation}
V_{\text{eff}}\left( E\right) =V_{a}+{|V_{am}|{ } ^{2}\over E-D{ }
 _{3D}} . \label{Veff}
\end{equation}
Let us at first reproduce, with some modifications, a renormalization
procedure for collisions in free space ($V_{\text{conf}}=0$) realized
 in Ref.\
\cite{K02}. The wavefunction in the momentum representation
\begin{equation}
\tilde{\psi }_{a}\left( {\bf q}\right) =\left( 2\pi \right) ^{-3
/2}\int d^{3}r \psi _{a}\left( {\bf r}\right) \exp\left( -i{\bf
 q}{\bf r}\right)
\end{equation}
can be represented as
\begin{equation}
\tilde{\psi }_{a}\left( {\bf q}\right) =\delta \left( {\bf q}-{\bf
 p}_{0}\right) +{m\over p^{2}_{0}-q^{2}+i0}\left( 2\pi \right)
 ^{-3}T_{\text{free}}\left( p_{0}\right)  ,
\end{equation}
where $p_{0}=\sqrt{mE}$  is the collision momentum and the $T$ matrix
 $T_{\text{free}}\left( p_{0}\right) $
obeys the Lippmann-Schwinger equation
\begin{equation}
T_{\text{free}}\left( p_{0}\right) =V_{\text{eff}}\left( E\right)
 \left\lbrack 1+{1\over 2\pi { } ^{2}}T_{\text{free}}\left(
 p_{0}\right) \int\limits^{p{ } _{c}}_{0}{m q^{2}dq\over
 p^{2}_{0}-q^{2}+i0}\right\rbrack  . \label{LS3D}
\end{equation}
The $\delta $ function potential in the coordinate representation
 leads to a
constant potential in the momentum representation and, therefore, to a
divergent integral in a Lippmann-Schwinger equation. This integral is
regularized in Eq.\ (\ref{LS3D}) by introduction of a momentum cutoff
 $p_{c}$.
The zero-energy limit of $T_{\text{free}}\left( p_{0}\right) $ should
 reproduce the dependence of the
elastic scattering length on the external magnetic field $B$ (see
 Ref.\
\cite{TTHK99})
\begin{equation}
T_{\text{free}}\left( 0\right) ={4\pi \over m}a_{3D}\left( 1+{\Delta
 \over B_{0}-B}\right) ,
\end{equation}
where $a_{3D}$  is the background elastic scattering length, $B_{0}$
 is the
resonant value of the magnetic field, and $\Delta $ is the
 phenomenological
resonance strength. The bound state energy $D_{3D}$  is proportional
 to the
magnetic field, $D_{3D}=\mu B+$const, where $\mu $ is the difference
 between the
magnetic momenta of an atomic pair in the open and closed channels.
 As a
result the non-renormalized parameters in Eq.\ (\ref{CC3D}) can be
 related
to $a_{3D}$, $B_{0}$, $\Delta $, and $\mu $ as
\begin{eqnarray}
&&V_{a}={4\pi \over m}a_{3D}\left( 1-{2\over \pi }a_{3D}p_{c}\right)
 ^{-1} \nonumber
\\
&&|V_{am}|^{2}={4\pi \over m}a_{3D}\mu \Delta \left( 1-{2\over \pi
 }a_{3D}p_{c}\right) ^{-2} \label{renorm}
\\
&&D_{3D}=\mu \left\lbrack B-B_{0}-\Delta +\Delta \left( 1-{2\over \pi
 }a_{3D}p_{c}\right) ^{-1}\right\rbrack  , \nonumber
\end{eqnarray}
reproducing result of Ref.\ \cite{K02}. In the limit
 $p_{c}\rightarrow \infty $ the $T$ matrix
can be expressed as
\begin{eqnarray}
&&T_{\text{free}}\left( p_{0}\right) =-i{4\pi \over m} a_{3D}
 \nonumber
\\
&&\times {p^{2}_{0}-mD{ } _{3D}\over
 a_{3D}p^{3}_{0}-ip^{2}_{0}-ma_{3D}D_{3D}p_{0}+i m\left( D_{3D}+\mu
 \Delta \right) } . \label{Tfree}
\end{eqnarray}
Consider now a case of cylindrical harmonic confinement with the
transverse frequency $\omega _{\perp }$,
\begin{equation}
V_{\text{conf}}={m\over 4}\omega ^{2}_{\perp }\rho ^{2} ,
\end{equation}
where $\rho ^{2}=x^{2}+y^{2}$  and $x$, $y$, and $z$ are the
 components of a vector ${\bf r}$. The
Schroedinger equation (\ref{Eff3D}) can be written out as
\begin{equation}
E\psi _{a}\left( {\bf r}\right) =\left\lbrack -{1\over m}{\partial {
 } ^{2}\over \partial z{ } ^{2}}+\hat{H}_{\perp }+V_{\text{eff}}\left
( E\right) \delta \left( {\bf r}\right) \right\rbrack \psi _{a}\left(
 {\bf r}\right)  ,
\end{equation}
where
\begin{equation}
\hat{H}_{\perp }=-{1\over m}\left( {\partial { } ^{2}\over \partial
 \rho { } ^{2}}+{1\over \rho }{\partial \over \partial \rho }+{1\over
 \rho { } ^{2}} {\partial { } ^{2}\over \partial \theta { }
 ^{2}}\right) +V_{\text{conf}}\left( \rho \right)
\end{equation}
is the transverse Hamiltonian and $\theta $ is the angular coordinate
 in the
$xy$ plane.

The eigenstates of $\hat{H}_{\perp }$, denoted as $|Nm_{z}\rangle $
(see Ref.\ \cite{MBO04}),
satisfy
\begin{equation}
\hat{H}_{\perp }|Nm_{z}\rangle =\left( N+|m_{z}|+1\right) \omega
 _{\perp }|Nm_{z}\rangle ,
\end{equation}
where $m_{z}$  is the angular momentum. The eigenfunctions can be
represented in terms of generalized Lagguere polynomials. The value of
$|Nm_{z}\rangle $ at the origin is zero for odd $N$ and independent
 of $N$ for even $N=2n$,
\begin{equation}
\langle 0|2nm_{z}\rangle ={1\over \sqrt{\pi }a{ } _{\perp }}\delta
 _{0m_{z}} ,
\end{equation}
where
\begin{equation}
a_{\perp }=\sqrt{{2\over m\omega { } _{\perp }}}
\end{equation}
is the transverse harmonic oscillator length. Therefore matrix
elements of the interatomic interaction
\begin{equation}
\langle 2n^\prime m^\prime _{z}|V_{\text{eff}}\delta \left( {\bf
 r}\right) |2n,m_{z}\rangle ={1\over \pi a{ } ^{2}_{\perp
 }}V_{\text{eff}}\delta _{0m_{z}}\delta _{0m^\prime _z}\delta _{z}
\end{equation}
are independent of $n$ and $n^\prime $.

Let us represent the wavefunction $\psi _{a}\left( {\bf r}\right) $ as
\begin{equation}
\psi _{a}\left( {\bf r}\right) =\left( 2\pi \right) ^{-1
/2}\sum\limits^{\infty }_{n=0}\int\limits^{\infty }_{-\infty }dq
 \tilde{\psi }_{n}\left( q\right)  e^{iqz}|2n0\rangle  .
\end{equation}
(The components proportional to $|Nm_{z}\rangle $ with odd $N$ or
 $m_{z}\neq 0$ are uncoupled
and, therefore, could be excluded.) The coefficients $\tilde{\psi
 }_{n}\left( q\right) $ satisfy the set
of coupled equations
\begin{eqnarray}
E\tilde{\psi }_{n}\left( q\right) =\left\lbrack {q{ } ^{2}\over m}
+\left( 2n+1\right) \omega _{\perp }\right\rbrack \tilde{\psi
 }_{n}\left( q\right)  \nonumber
\\
+{1\over 2\pi ^{2}a{ } ^{2}_{\perp }}V_{\text{eff}}\left( E\right)
 \sum\limits^{\infty }_{n^\prime =0}\int\limits^{\infty }_{-\infty
 }dq^\prime \tilde{\psi }_{n^\prime }\left( q^\prime \right)  .
\end{eqnarray}
For a collision of two atoms in a transverse state $n$ with a relative
axial momentum $p_{n}=\sqrt{m\left\lbrack E-\left( 2n+1\right) \omega
 _{\perp }\right\rbrack }$ the coefficients can be expressed as
\begin{equation}
\tilde{\psi }_{n^\prime }\left( q\right) =\delta \left( q-p_{n}\right
) \delta _{n^\prime n}+{m\over p^{2}_{n^\prime }-q^{2}+i0}{1\over
 2\pi }T_{n^\prime n}\left( p_{0}\right)  .
\end{equation}
The elements of $T$ matrix $T_{n^\prime n}\left( p_{0}\right) $
 satisfy the Lippmann-Schwinger
equations
\begin{equation}
T_{n^\prime n}\left( p_{0}\right) ={1\over \pi a{ } ^{2}_{\perp
 }}V_{\text{eff}}\left( E\right) \left\lbrack 1-{i\over
 2}m\sum\limits^{n{ } _{c}}_{n^{\prime\prime}=0}{1\over p{ }
 _{n^{\prime\prime}}}T_{n^{\prime\prime}n}\left( p_{0}\right)
 \right\rbrack  , \label{LSconf}
\end{equation}
where the level cutoff $n_{c}$  regularizes the divergent series. The
right-hand-side of Eq.\ (\ref{LSconf}) is independent of $n$ and
 $n^\prime $.
Therefore $T_{n^\prime n}\left( p_{0}\right) $ is independent of $n$
 and $n^\prime $  as well and has a form,
\begin{eqnarray}
&&T_{n^\prime n}\left( p_{0}\right) =T_{\text{conf}}\left(
 p_{0}\right)  \nonumber
\\
&&=V_{\text{eff}}\left( E\right) \left\lbrack \pi a^{2}_{\perp }
+{i\over 2}V_{\text{eff}}\left( E\right) m\sum\limits^{n{ }
 _{c}}_{n^{\prime\prime}=0}{1\over p{ }
 _{n^{\prime\prime}}}\right\rbrack ^{-1} .
\end{eqnarray}
A substitution of Eqs.\ (\ref{Veff}), (\ref{renorm}), and
(\ref{Tfree}) allows us to express $T_{\text{conf}}\left( p_{0}\right
) $ in terms of the physical
parameters as,
\begin{eqnarray}
&&T_{\text{conf}}\left( p_{0}\right) ={4\over ma{ } _{\perp
 }}\Biggl\{{4\pi a{ } _{\perp }\over mT_{\text{free}}\left(
 p_{0}\right) } \nonumber
\\
&&+\sum\limits^{n{ } _{c}}_{n=0}\left\lbrack n-\left( p_{0}a_{\perp }
/2\right) ^{2}\right\rbrack ^{-1/2}-{2\over \pi }p_{c}a_{\perp
 }\Biggr\}^{-1} .
\end{eqnarray}
This expression has a finite limit at $n_{c}\rightarrow \infty ,
 p_{c}\rightarrow \infty $  for
$a_{\perp }p_{c}=\pi \sqrt{n_{c}-\left( p_{0}a_{\perp }/2\right) { }
 ^{2}}$. This limit can be written out as,
\begin{equation}
T_{\text{conf}}\left( p_{0}\right) ={4\over ma{ } _{\perp
 }}\left\lbrack {4\pi a{ } _{\perp }\over mT_{\text{free}}\left(
 p_{0}\right) }+\zeta \left( {1\over 2},-\left( {a_{\perp }p{ }
 _{0}\over 2}\right) ^{2}\right) \right\rbrack ^{-1} , \label{Tconf}
\end{equation}
where
\begin{equation}
\zeta \left( {1\over 2},\alpha \right) =\mathrel{\mathop
 \mathrm{\lim}_{n_{c}\rightarrow \infty }}\left\lbrack
 \sum\limits^{n{ } _{c}}_{n=0}\left( n+\alpha \right) ^{-1/2}-2\left(
 n_{c}+\alpha \right) ^{1/2}\right\rbrack  , \label{zeta}
\end{equation}
with $-2\pi <\arg\left( n+\alpha \right) \le 0$, is the Hurwitz zeta
 function (see Refs.\
\cite{MBO04,BE53}). Equation (\ref{Tconf}) is similar to Eq.\ (10) in
Ref.\ \cite{BMO03}, but the elastic scattering length is replaced by
 an
energy-dependent function ${m\over 4\pi }T_{\text{free}}\left(
 p_{0}\right) $. For a case of non-resonant
scattering ($\Delta =0$) Eq.\ (\ref{Tconf}) reproduces the results of
 Refs.\
\cite{BMO03,MBO04}. This derivation can also be implemented for a case
of three-dimensional harmonic confinement, giving some justification
 to
the use of an energy-dependent resonant scattering length in Ref.\
\cite{BTJ02,BTJ03}, although a Fermi pseudopotential is used there
rather then $\delta $ function.

\section{Feshbach resonance scattering in one dimension
\label{sec1D}}

\subsection{Atom-molecule model}

Let us treat a bound state of an atomic pair in the closed channel
as a molecule. A many-body one-dimensional system of coupled atoms and
two-atom molecules can be described by the Hamiltonian similar to one
used in a related three-dimensional problem (see Refs.\
\cite{TTHK99,YBJW99}),
\begin{eqnarray}
\hat{H}_{am}&=&\int\limits^{\infty }_{-\infty }d x \biggl\{\hat{\Psi
 }^{\dag }_{m}\left( x\right) \left( -{1\over 4m}{d{ } ^{2}\over dx{
 } ^{2}}+D_{1D}\right) \hat{\Psi }_{m}\left( x\right) + \nonumber
\\
&&+\hat{\Psi }^{\dag }_{a}\left( x\right) \left\lbrack -{1\over 2m}
 {d{ } ^{2}\over dx{ } ^{2}}+{U{ } _{a}\over 2}\hat{\Psi }^{\dag
 }_{a}\left( x\right) \hat{\Psi }_{a}\left( x\right) \right\rbrack
 \hat{\Psi }_{a}\left( x\right)  + \nonumber
\\
&&+ \left\lbrack g \hat{\Psi }^{\dag }_{m}\left( x\right) \hat{\Psi
 }_{a}\left( x\right) \hat{\Psi }_{a}\left( x\right)  +
 \mathrm{h.c.}\right\rbrack \biggr \} . \label{Ham}
\end{eqnarray}
Here $\hat{\Psi }_{a}\left( y\right) $ and $\hat{\Psi }_{m}\left(
 x\right) $ are the annihilation operators for the
atomic and molecular fields, respectively, $U_{a}$  is the strength
 of an
interatomic interaction, and $D_{1D}$  is the energy of the molecular
 state
counted from the open channel threshold.

A state vector of the two-atom system can be represented as a
superposition of atomic and molecular states,
\begin{eqnarray}
&&|\Psi ^{am}_{2}\rangle =e^{iPx}\Bigl\lbrack \varphi
 ^{am}_{1}\hat{\Psi }^{\dag }_{m}\left( x\right)  \nonumber
\\
&&+{1\over \sqrt{2}}\int\limits^{\infty }_{-\infty }dy \varphi
 _{0}\left( y\right) \hat{\Psi }^{\dag }_{a}\left( x-{y\over 2}\right
) \hat{\Psi }^{\dag }_{a}\left( x+{y\over 2}\right) \Bigr\rbrack
|0\rangle  , \label{Psiam2}
\end{eqnarray}
where $P$ is a center-of-mass momentum, $y$ is an interatomic
distance, and $|0\rangle $ is the physical vacuum state.

A substitution of Eq.\ (\ref{Psiam2}) into the Schroedinger
equation
\begin{equation}
\left( {P{ } ^{2}\over 4m}+E\right) |\Psi ^{am}_{2}\rangle
 =\hat{H}_{am}|\Psi ^{am}_{2}\rangle  \label{Scham}
\end{equation}
gives the coupled equations for the coefficients $\varphi _{0}\left(
 y\right) $ and $\varphi ^{am}_{1}$
\begin{eqnarray}
&&E\varphi _{0}\left( y\right) =-{1\over m} {d^{2}\varphi { }
 _{0}\over dy{ } ^{2}}+\left\lbrack U_{a}\varphi _{0}\left( 0\right)
+\sqrt{2}g^{*}\varphi ^{am}_{1}\right\rbrack \delta \left( y\right)
 \nonumber
\\
&&{}\label{CC1D}
\\
&&E\varphi ^{am}_{1}=D_{1D}\varphi ^{am}_{1}+\sqrt{2}g\varphi
 _{0}\left( 0\right)  \nonumber
\end{eqnarray}
Elimination of $\varphi ^{am}_{1}$  from these equations leads to a
 single
Schroedinger equation
\begin{equation}
E\varphi _{0}\left( y\right) =-{1\over m} {d^{2}\varphi { } _{0}\over
 dy{ } ^{2}}+U_{\text{eff}}\left( E\right) \delta \left( y\right)
 \varphi _{0}\left( 0\right)  \label{Effam}
\end{equation}
with the energy-dependent effective interaction strength
\begin{equation}
U_{\text{eff}}\left( E\right) =U_{a}+{2|g|{ } ^{2}\over E-D{ } _{1D}}
 . \label{Ueffam}
\end{equation}
\subsection{Two-state model}

An alternative description of Feshbach resonances in
one-dimensional systems involves atoms with two internal
states, $a$ and $b$, associated to the annihilation operators
$\hat{\Psi }_{a}\left( x\right) $ and $\hat{\Psi }_{b}\left( x\right)
 $, respectively. The many-body Hamiltonian of
this system has the form,
\begin{eqnarray}
&&\hat{H}_{ab}=\int\limits^{\infty }_{-\infty }d x \biggl\{-{1\over
 2m}\sum\limits^{}_{\alpha =a,b}\hat{\Psi }^{\dag }_{\alpha }\left(
 x\right) {d{ } ^{2}\over dx{ } ^{2}}\hat{\Psi }_{\alpha }\left(
 x\right)  \nonumber
\\
&&+D_{b}\hat{\Psi }^{\dag }_{b}\left( x\right) \hat{\Psi }_{b}\left(
 x\right)  +{U{ } _{a}\over 2}\hat{\Psi }^{\dag }_{a}\left( x\right)
 \hat{\Psi }^{\dag }_{a}\left( x\right) \hat{\Psi }_{a}\left( x\right
) \hat{\Psi }_{a}\left( x\right)  \nonumber
\\
&&+ U_{b}\hat{\Psi }^{\dag }_{a}\left( x\right) \hat{\Psi }^{\dag
 }_{b}\left( x\right) \hat{\Psi }_{a}\left( x\right) \hat{\Psi
 }_{b}\left( x\right)  \nonumber
\\
&&+ \left\lbrack {U{ } _{ab}\over \sqrt{2}} \hat{\Psi }^{\dag
 }_{a}\left( x\right) \hat{\Psi }^{\dag }_{b}\left( x\right)
 \hat{\Psi }_{a}\left( x\right) \hat{\Psi }_{a}\left( x\right)  +
 \mathrm{h.c.}\right\rbrack \biggr \} , \label{Hab}
\end{eqnarray}
where $D_{b}$  is the energy mismatch between the states $a$ and $b$
 and
$U_{ab}$  describes transitions between the states on atomic
 collisions. A
state vector of two-atom system can be represented as a superposition
of different atomic states
\begin{eqnarray}
|\Psi ^{ab}_{2}\rangle &=&e^{iPx}\int\limits^{\infty }_{^{-\infty
 }}dy\biggl\lbrack \varphi ^{ab}_{1}\left( y\right) \hat{\Psi }^{\dag
 }_{a}\left( x-{y\over 2}\right) \hat{\Psi }^{\dag }_{b}\left( x
+{y\over 2}\right)  \nonumber
\\
&&+ {1\over \sqrt{2}}\varphi _{0}\left( y\right) \hat{\Psi }^{\dag
 }_{a}\left( x-{y\over 2}\right) \hat{\Psi }^{\dag }_{a}\left( x
+{y\over 2}\right) \biggr\rbrack |0\rangle  , \label{Psiab2}
\end{eqnarray}
where the coefficients $\varphi _{0}\left( y\right) $ and $\varphi
 ^{ab}_{1}\left( y\right) $  satisfy the set of
coupled equations
\begin{eqnarray}
E\varphi _{0}\left( y\right) &=&-{1\over m} {d^{2}\varphi { }
 _{0}\over dy{ } ^{2}}+\Bigl\lbrack U_{a}\varphi _{0}\left( 0\right)
 \nonumber
\\
&&+U^{*}_{ab}\varphi ^{ab}_{1}\left( 0\right) \Bigr\rbrack \delta
 \left( y\right)  \label{CCab0}
\\
E\varphi ^{ab}_{1}\left( y\right) &=&-{1\over m} {d^{2}\varphi { }
 ^{ab}_{1}\over dy{ } ^{2}}+D_{1D}\varphi ^{ab}_{1}\left( y\right)
 \nonumber
\\
&&+\left\lbrack U_{ab}\varphi _{0}\left( 0\right) +U_{b}\varphi
 ^{ab}_{1}\left( 0\right) \right\rbrack \delta \left( y\right)
 \label{CCab1}
\end{eqnarray}
A Feshbach resonance can appear if one of the channels, $ab$,
related to the coefficient $\varphi ^{ab}_{1}\left( y\right) $, has a
 bound state. Such state can
appear when $U_{b}<0$ and has the energy $D_{1D}=D_{b}-{m\over
 4}U^{2}_{b}$. In following $D_{b}$  and
$|U_{b}|$ will be tended to infinity keeping a fixed value of
 $D_{1D}$. This
limit allows us to neglect the third channel, $bb$, which will be
infinitely distanced from the other ones.

The coefficient $\varphi ^{ab}_{1}\left( y\right) $ can be expressed
 from Eq.\ (\ref{CCab1}) as
\begin{equation}
\varphi ^{ab}_{1}\left( y\right) =- {U{ } _{ab}\over 2\kappa _{b}/m
+U{ } _{b}}\varphi _{0}\left( 0\right) \exp\left( -\kappa _{b}y\right
)  , \label{Phiab1}
\end{equation}
where $\kappa _{b}=\sqrt{m\left( D_{b}-E\right) }$.

A substitution of Eq.\ (\ref{Phiab1}) into Eq.\ (\ref{CCab0})
leads again to the one-channel equation (\ref{Effam}), but now the
effective strength is expressed as
\begin{equation}
U_{\text{eff}}\left( E\right) =U_{a}-{|U_{ab}|{ } ^{2}\over 2\kappa
 _{b}/m+U{ } _{b}} . \label{Ueffab}
\end{equation}
This expression tends to Eq.\ (\ref{Ueffam}) in the limit
 $U_{b}\rightarrow -\infty $,
while
\begin{equation}
D_{b}=D_{1D}+{m\over 4}U^{2}_{b},\qquad U_{ab}={2\over \sqrt{m|U_{b}
|}}g .
\end{equation}
\subsection{$T$ matrix}

Equation (\ref{Effam}) has a simple exact solution. The
wavefunction in the momentum representation can be expressed as
\begin{eqnarray}
\tilde{\varphi }_{0}\left( q\right) =\left( 2\pi \right) ^{-1
/2}\int\limits^{\infty }_{-\infty }dy \psi _{0}\left( y\right)
 \exp\left( -i q y\right)  \nonumber
\\
=\delta \left( p_{0}-q\right) +{m\over p^{2}_{0}-q^{2}+i0} {1\over
 2\pi }T_{1D}\left( p_{0}\right)
\end{eqnarray}
where $p_{0}=\sqrt{mE}$ is the collision momentum and
the one-dimensional $T$ matrix satisfy the
Lippmann-Schwinger equation,
\begin{equation}
T_{1D}\left( p_{0}\right) =U_{\text{eff}}\left( E\right) \left\lbrack
 1+{1\over 2\pi }T_{1D}\left( p_{0}\right) \int\limits^{\infty
 }_{-\infty }{m dq\over p^{2}_{0}-q^{2}+i0}\right\rbrack .
\end{equation}
Its solution has the form
\begin{equation}
T_{1D}\left( p_{0}\right) =U_{\text{eff}}\left( E\right) \left\lbrack
 1+{i m\over 2p{ } _{0}}U_{\text{eff}}\left( E\right) \right\rbrack
 ^{-1}. \label{T1D}
\end{equation}
\section{Results and discussion\label{secRes}}

\subsection{One-dimensional parameters}

The limit of small collision momentum $p_{0}$  corresponds to small
values of the second argument of the Hurwitz zeta function in Eq.\
(\ref{Tconf}). A substitution of expansion
\begin{equation}
\zeta \left( {1\over 2},\alpha \right) \mathrel{\mathop \sim_{\alpha
 \rightarrow 0}}{1\over \sqrt{\alpha }}-C+O\left( \alpha \right) ,
 \sqrt{-|\alpha |}=-i\sqrt{|\alpha |}, \label{alpha0}
\end{equation}
where $C\approx 1.4603$ is the Olshanii constant (see Ref.\
 \cite{O98}),
into Eq.\ (\ref{Tconf}) leads to an expression which coincides with
 Eq.\
(\ref{T1D}) if
\begin{equation}
U_{\text{eff}}\left( E\right) ={1\over \pi a{ } ^{2}_{\perp
 }}T_{\text{free}}\left( E\right) \left\lbrack 1-{C m\over 4\pi a{ }
 _{\perp }}T_{\text{free}}\left( E\right) \right\rbrack ^{-1}.
 \label{Ueff}
\end{equation}
This expression is nothing else but the Olshanii formula
(see Ref.\ \cite{O98}) where the elastic scattering length and
the one-dimensional interaction strength are replaced by
${m\over 4\pi }T_{\text{free}}\left( p_{0}\right) $ and the
 energy-dependent strength $U_{\text{eff}}\left( E\right) $,
respectively.

Equation (\ref{Ueff}), in a combination with Eqs.\ (\ref{Tfree})
and (\ref{Ueffam}), allows to relate the one-dimensional parameters
 $U_{a}$,
$D_{1D}$, and $g$ to the physical parameters $a_{3D}$, $\Delta $,
 $B_{0}$, and $\mu $ as
\begin{eqnarray}
&&U_{a}={4a{ } _{3D}\over ma{ } ^{2}_{\perp }}\left( 1-C{a{ }
 _{3D}\over a{ } _{\perp }}\right) ^{-1} \label{U1D}
\\
&&|g|^{2}={2a_{3D}\mu \Delta \over ma{ } ^{2}_{\perp }}\left( 1-C{a{
 } _{3D}\over a{ } _{\perp }}\right) ^{-2} \label{g1D}
\\
&&D_{1D}=\mu \left( B-B_{0}\right) -\omega _{\perp }+C{a{ }
 _{3D}\over a{ } _{\perp }}\mu \Delta \left( 1-C{a{ } _{3D}\over a{ }
 _{\perp }}\right) ^{-1} . \label{D1D}
\end{eqnarray}
The confinement-induced resonance (see Refs.\ \cite{O98,BMO03})
not only scales here the interaction parameters, but also shifts the
Feshbach resonance by the last term in Eq.\ (\ref{D1D}). The term
 $\omega _{\perp }$
in Eq.\ (\ref{D1D}) reflects the shift of the continuum threshold due
to transverse oscillations.

\subsection{Scattering}

Let us introduce dimensionless variables (the scattering momentum
$k$, the elastic scattering length $a$, the detuning $b$, and the
 resonance
strength $d$) as,
\begin{eqnarray}
k&=&{p_{0}a{ } _{\perp }\over 2}=\sqrt{{E\over 2\omega { } _{\perp
 }}-{1\over 2}}\approx 8.1\times 10^{-3}\sqrt{{E\text{(nK)}\over
 \omega _{\perp }\text{(Mhz)}} - {1\over 2}} \nonumber
\\
a&=&{a{ } _{3D}\over a{ } _{\perp }}\approx 2.8\times
 10^{-3}\sqrt{m\text{(AMU)}\omega _{\perp }\text{(MHz)}}a_{3D}\text{
(nm)} \nonumber
\\
b&=&\mu {B-B{ } _{0}\over 2\omega { } _{\perp }}-{1\over 2}\approx
 4.4\mu \left( \mu _{B}\right) {\left( B-B_{0}\right) \text{(G)}\over
 \omega _{\perp }\text{(MHz)}}-{1\over 2} \label{dimless}
\\
d&=&{a_{3D}\mu \Delta \over 2a_{\perp }\omega { } _{\perp }} \nonumber
\\
&&\approx 1.2\times 10^{-2}{a_{3d}\text{(nm)}\mu \left( \mu
 _{B}\right) \Delta \text{(G)}\sqrt{m\text{(AMU)}}\over \sqrt{\omega
 _{\perp }\text{(MHz)}}} . \nonumber
\end{eqnarray}
For example, for $\omega _{\perp }=2\pi \times 80$ Khz, we have
 $a\approx 2.3\times 10^{-3}$,
$b\approx 64\left\lbrack B(\mathrm{G})-853\right\rbrack $, and
 $d\approx 0.14$ for the case of the 853 G
resonance in Na; $a\approx 0.07$, $b\approx 48\left\lbrack B
(\mathrm{G})-1007\right\rbrack $, and $d\approx 0.6$ for the
case of the 1007 G resonance in $^{87}$Rb; and $a\approx 0.31$,
$b\approx 39\left\lbrack B(\mathrm{G})-155\right\rbrack $, and
 $d\approx 130$ for the case of the resonance in
$^{85}$Rb.

The $T$ matrix for the scattering under cylindrical harmonic
confinement (\ref{Tconf}) is expressed in terms of these dimensionless
variables as
\begin{equation}
T_{\text{conf}}={4\over m\omega { } _{\perp }}{a k^{2}-a b+d\over
 \left\lbrack a k^{2}-a b+d\right\rbrack \zeta \left( 1
/2,-k^{2}\right) +k^{2}-b} . \label{TconfU}
\end{equation}
For small values of $k$ the zeta function can be approximated by
the expansion Eq.\ (\ref{alpha0}), leading to the expression for
one-dimensional $T$ matrix [cf. Eq.\ (\ref{T1D})]
\begin{eqnarray}
&&T_{1D}={4\over m\omega { } _{\perp }} \nonumber
\\
&&\times {a k^{3}-\left( a b+d\right) k\over \left( 1-Ca\right) k^{3}
+i a k^{2}-\left\lbrack b\left( 1-Ca\right) +Cd\right\rbrack k-i\left
( ab-d\right) } . \nonumber
\\
{}\label{T1DU}
\end{eqnarray}
A study of the opposite, high $k$, limit requires an expansion for
$\zeta \left( {1\over 2},\alpha \right) $ at high values of $\alpha
 $. For $\alpha >0$ such expansion can be obtained
from the integral representation (1.10.7) in Ref.\ \cite{BE53}
\begin{equation}
\zeta \left( {1\over 2},\alpha \right) \mathrel{\mathop \sim_{\alpha
 \rightarrow \infty }}-2\alpha ^{1/2}+{1\over 2}\alpha ^{-1/2}
+{1\over 24}\alpha ^{-3/2} . \label{alphap}
\end{equation}
At $\alpha <0$ the function $\zeta \left( {1\over 2},\alpha \right) $
 has complex values with the imaginary
part represented by a finite sum (see also Ref.\ \cite{MBO04}),
\begin{equation}
\zeta \left( {1\over 2},\alpha \right) =\zeta \left( {1\over
 2},\left\lbrack |\alpha |\right\rbrack +1-|\alpha |\right)
+i\sum\limits^{\left\lbrack |\alpha |\right\rbrack }_{n=0}\left(
|\alpha |-n\right) ^{-1/2},
\end{equation}
where $\left\lbrack |\alpha |\right\rbrack $ denotes the integer part
 of $|\alpha |$. In the limit $|\alpha |\rightarrow \infty $
this sum can be approximated by the Hurwith zeta function itself [see
Eq.\ (\ref{zeta})], leading to
\begin{eqnarray}
\zeta \left( {1\over 2},\alpha \right) \mathrel{\mathop \sim_{\alpha
 \rightarrow -\infty }}\zeta \left( {1\over 2},\left\lbrack |\alpha
|\right\rbrack +1-|\alpha |\right)  \nonumber
\\
+i\Biggl\{2\sqrt{|\alpha |}+\zeta \left( {1\over 2},|\alpha
|-\left\lbrack |\alpha |\right\rbrack \right) \Biggr\} .
 \label{alpham}
\end{eqnarray}
A substitution of the leading term of this expansion into Eq.\
(\ref{TconfU}) gives us
\begin{equation}
T_{3D}=-i{4\over m\omega { } _{\perp }}{a k^{2}-a b+d\over 2a
 k^{3}-ik^{2}-2\left( a b-d\right) k+ib} . \label{T3DU}
\end{equation}
This expression differs from $T_{\text{free}}$  (see Eq.\
(\ref{Tfree})) only
by a factor due to different definitions of $T$ matrix for the
scattering of the confined and free atoms.

The one-dimensional scattering can be characterized by reflection
probability
\begin{equation}
R=|f_{\text{even}}|^{2}=|{m\over 2p{ } _{0}}T_{\text{conf}}|^{2}
 \label{Refl}
\end{equation}
and by the phase of the scattering amplitude
\begin{equation}
\chi =\arg T_{\text{conf}}\left( k\right) -{\pi \over 2} .
 \label{Phase}
\end{equation}
\begin{figure}
\includegraphics[width=3.375in]{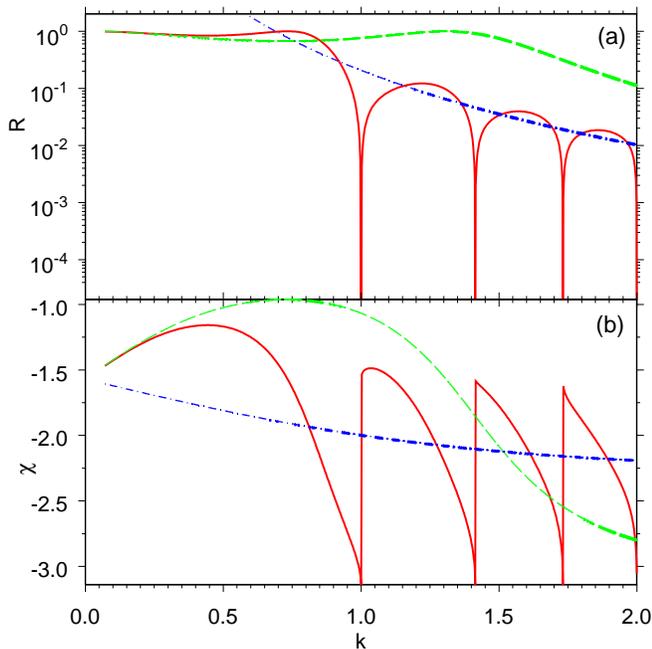}

\caption{The reflection probability [(a), see Eq.\
(\protect\ref{Refl})] and the scattering phase [(b), see Eq.\
(\protect\ref{Phase})] calculated as functions of the
dimensionless collision momentum $k$ [see Eq.\
(\protect\ref{dimless})] with the parameter values $a=0.1$, $b=0$, and
$d=1$. The solid, dashed, and dot-dashed lines are related,
respectively, to the exact expression  (\protect\ref{TconfU}), to
the one-dimensional approximation (\protect\ref{T1DU}), and to the
three-dimensional one (\protect\ref{T3DU}).} \label{FigSb0}

\end{figure}
\begin{figure}
\includegraphics[width=3.375in]{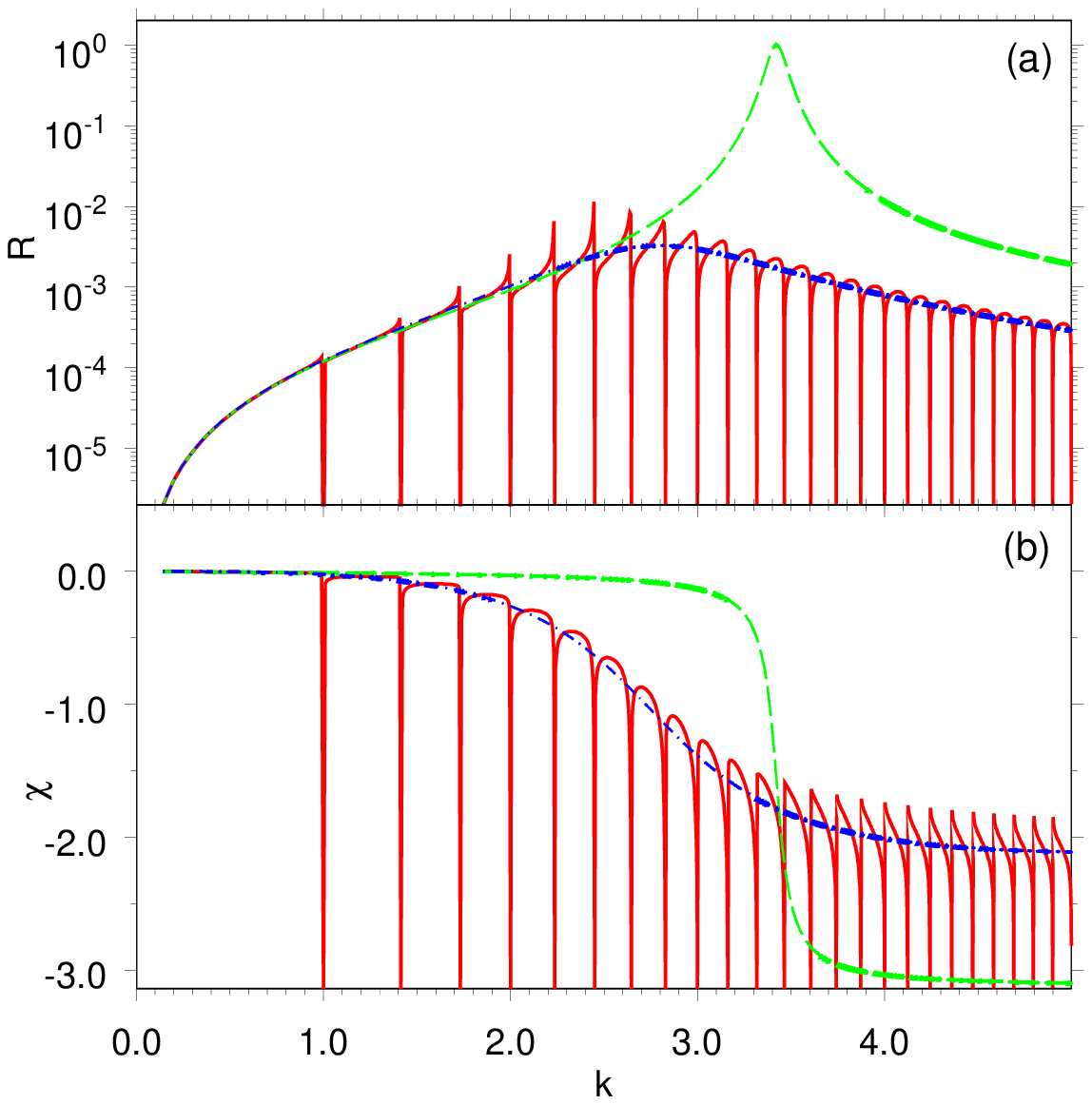}

\caption{The same as in Fig.\ \protect\ref{FigSb0}, but for the
parameter values $a=0.1$, $b=10$, and $d=1$.} \label{FigSb10}

\end{figure}
\begin{figure}
\includegraphics[width=3.375in]{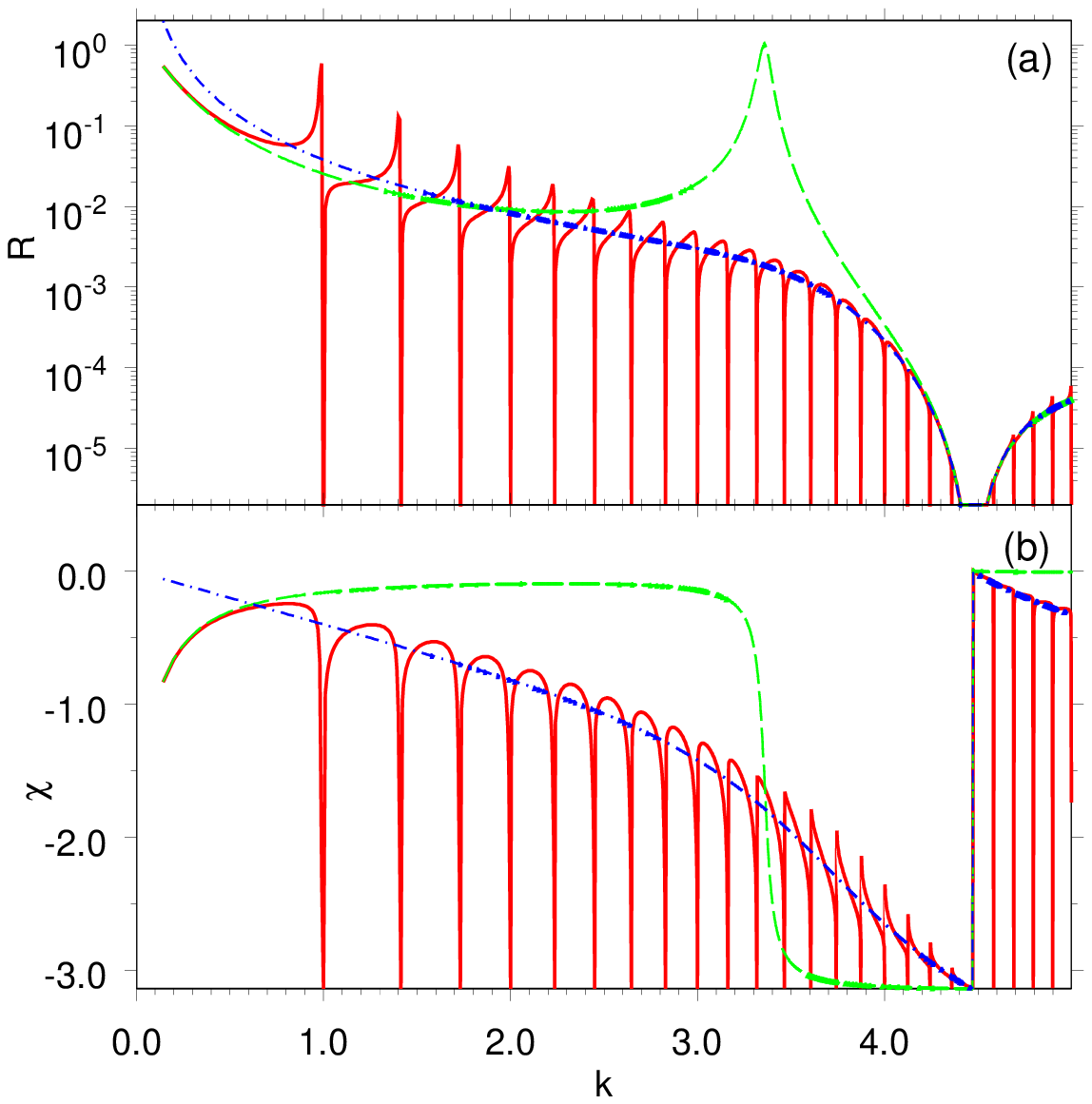}

\caption{The same as in Fig.\ \protect\ref{FigSb0}, but for the
parameter values $a=-0.1$, $b=10$, and $d=1$.} \label{FigSamb10}

\end{figure}

A comparison of results obtained with the exact expression
(\ref{TconfU}) and the approximate ones Eq.\ (\ref{T1DU}) and Eq.\
(\ref{T3DU}) is presented in Figs.\ \ref{FigSb0}, \ref{FigSb10},
and \ref{FigSamb10}. As one can see, the one-dimensional
approximation (\ref{T1DU}) is appropriate only at small values of
$k$. The three-dimensional approximation (\ref{T3DU}) reproduces an
average behavior at high $k$, but it cannot reproduce the jumps
appearing due to opening of transverse channels at integer values
of $k^{2}$.

\subsection{Bound states}

The bound state energies of two atoms in the atomic waveguide are
given by the poles of $T_{\text{conf}}\left( p_{0}\right) $ on the
 positive imaginary axis of $p_{0}$.
Equation (\ref{TconfU}) leads to the transcendent equation in $x=-ik$
\begin{equation}
{x^{2}+b\over ax^{2}+ab-d}=-\zeta \left( {1\over 2},x^{2}\right) , x
>0, \label{x}
\end{equation}
determining the dimensionless binding energy
\begin{equation}
\epsilon _{b}=x^{2} , \label{eb}
\end{equation}
such that the bound state energies are
\begin{equation}
E=-2\omega _{\perp }\left( \epsilon _{b}-{1\over 2}\right) .
\end{equation}
For shallow bound states ($|E-\omega _{\perp }|\ll \omega _{\perp }$,
 $x\ll 1$), a substitution of the
expansion (\ref{alpha0}) gives a cubic equation
\begin{equation}
\left( 1-Ca\right) x^{3}+ax^{2}+\left\lbrack b\left( 1-Ca\right)
+Cd\right\rbrack x+ab-d=0 \label{x1D}
\end{equation}
for bound states in the related one-dimensional system. A similar
equation has been used in Ref.\ \cite{KD98} for evaluation of binding
energies of ``bosonic mesons''.

For deep bound states  ($|E|\gg \omega _{\perp }$, $x\gg 1$) a
 substitution of the
leading term of the expansion (\ref{alphap}) into Eq.\ (\ref{x})
results in a cubic equation
\begin{equation}
2ax^{3}-x^{2}+2\left( ab-d\right) x-b=0 , \label{x3D}
\end{equation}
determined bound states of free particles (similar equation has
been considered in Refs.\ \cite{Greene,DK04}).

\begin{figure}
\includegraphics[width=3.375in]{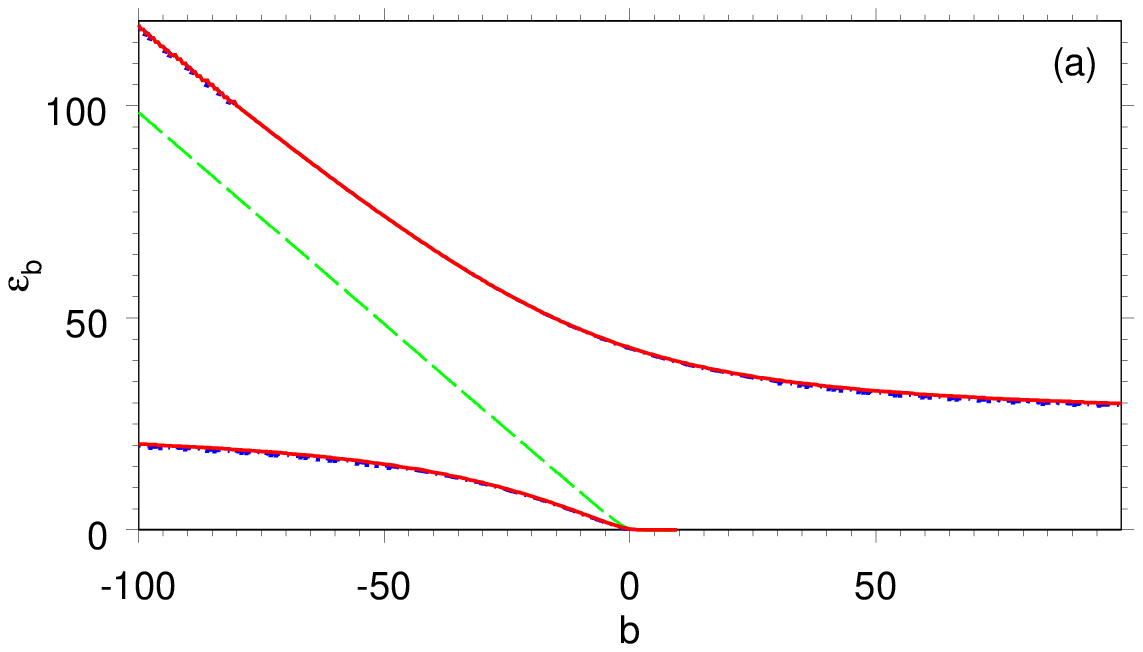}
\includegraphics[width=3.375in]{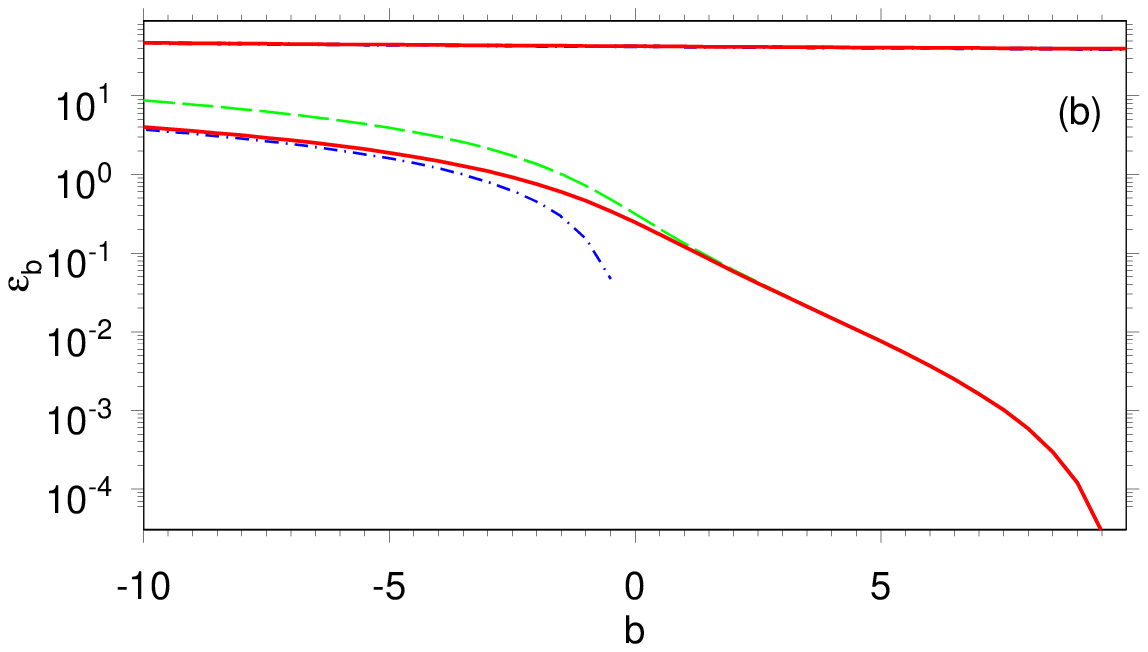}

\caption{The binding energy [see Eq.\ (\protect\ref{eb})]
calculated as functions of the dimensionless detuning $b$ [see Eq.\
(\protect\ref{dimless})] with the parameter values $a=0.1$ and $d=1$.
 The
solid, dashed, and dot-dashed lines are related, respectively, to the
exact expression  (\protect\ref{x}), to the one-dimensional
approximation (\protect\ref{x1D}), and to the three-dimensional one
(\protect\ref{x3D}). The liner plot (a) demonstrates a general
behavior, while the logarithmic one (b) highlights the near-resonance
region. The solid and dot-dashed lines almost coincide in the part
(a).} \label{FigBp}

\end{figure}

\begin{figure}
\includegraphics[width=3.375in]{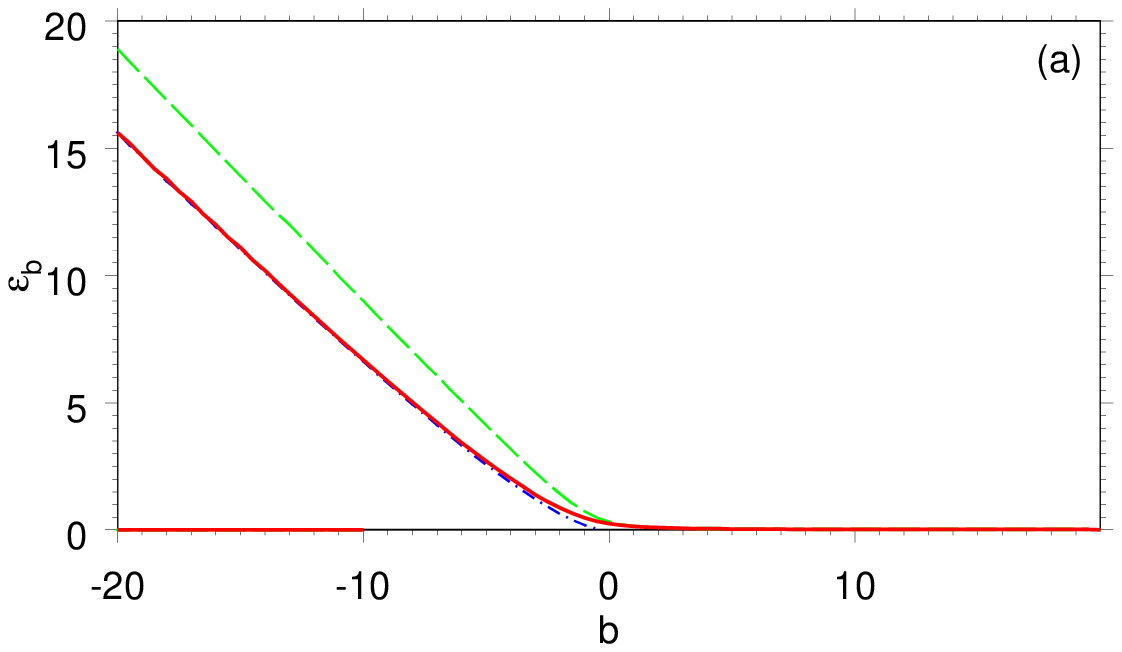}
\includegraphics[width=3.375in]{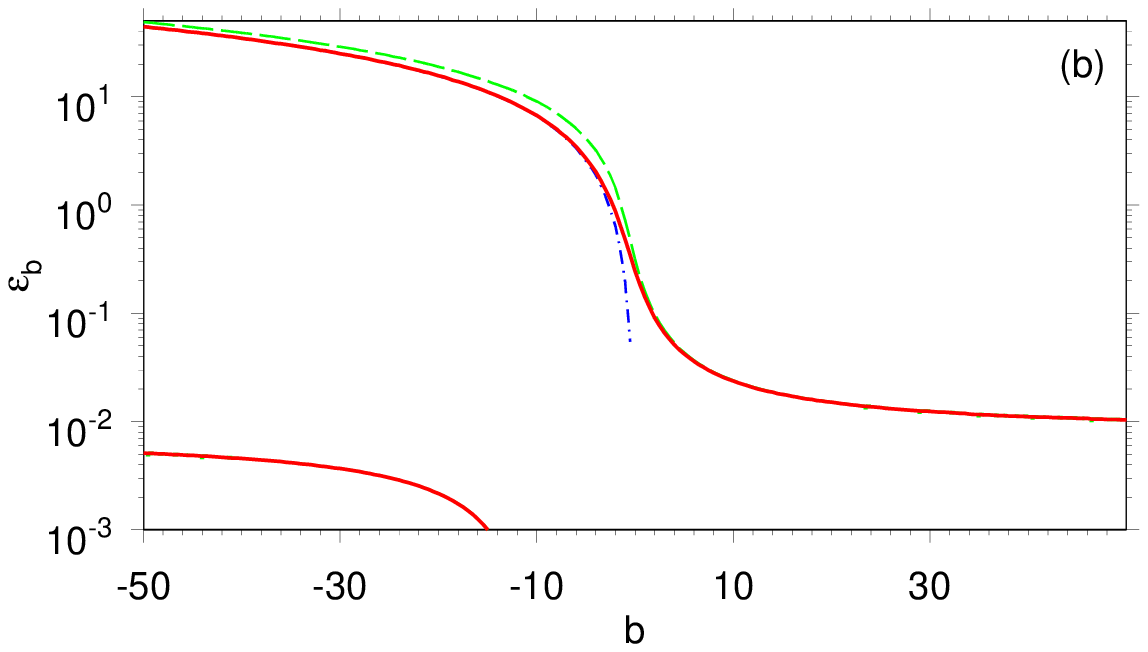}

\caption{The same as in Fig.\ \protect\ref{FigBp}, but for the
parameter values $a=-0.1$ and $d=1$. The lower solid and dashed lines
almost coincide in the part (b).} \label{FigBn}

\end{figure}

A general behavior of bound states can be seen from a qualitative
analysis of Eq.\ (\ref{x}). Its right hand side varies monotonically
from $-\infty $ to $\infty $ for $0<x<\infty $. The left hand side is
 a monotonic function of
$x$ for $x>0$ unless $b<d/a$. Therefore Eq.\ (\ref{x}) has two real
 positive
roots for $b<d/a$ and only one such root otherwise. The two roots
correspond to two bound states. One of them tends at $b\rightarrow
 -\infty $ to the bound
state in the closed channel. At $b\rightarrow \infty $ it tends to
 the bound state of the
related one-channel system analyzed in Ref.\ \cite{BMO03}. The second
bound state tends to the state of the one-channel system at
 $b\rightarrow \infty $ and
vanishes at the continuum threshold at $b=d/a$. A numerical solution
 of
Eq.\ (\ref{x}) (see Figs.\ \ref{FigBp} and \ref{FigBn}) confirms the
results of the quantitative analysis, demonstrating an intermediate
behavior of the confined problem between the one-dimensional and
three-dimensional approximations. In a case of a positive background
scattering length (see Fig.\ \ref{FigBp}) the bound states of the
confined problem tend at $|b|\rightarrow \infty $ to the bound states
 of the three-
dimensional model, while the one-dimensional one has a single bound
state only. An opposite situation takes place at negative scattering
length (see Fig.\ \ref{FigBn}), when the three-dimensional model has
one bound state only, and the bound states of the confined problem
tend at $|b|\rightarrow \infty $ to the two bound states of the
 one-dimensional model.


\section*{Conclusions}

A problem of two-channel scattering of atoms under cylindrical
harmonic confinement can be solved using a renormalization procedure.

Many-body problems involving two-channel scattering can be
described in one dimension by two models: the atom-molecule one and
the two-state one. Parameters of these models can be expressed in
terms of three-dimensional scattering parameters, using the
solution of the confined problem. Scattering amplitudes and bound
states of the confined system incorporate both proprieties of the
related one-dimensional and three-dimensional systems, as well as
specific peculiarities.

\end{document}